\documentclass[longauth,traditabstract,letter]{aa}
\usepackage{longtable}
\usepackage{natbib,afterpage}
\usepackage{graphicx,color}
\usepackage{amssymb,times,graphics, subfigure}
\usepackage[english]{babel}
\usepackage{txfonts}

\def\Msun{\hbox{$M_{\odot}$}}               %% solar mass
\def\Rstar{\hbox{$R_{\star}$}}              %% stellar radius
\def\Mdot{\hbox{$\dot{M}$}}               %% Mdot      
\def\Teff{\hbox{$\rm{T}_{\rm eff}$}}            %% T_eff
\def\arcsec{\hbox{$^{\prime\prime}$}}
\def\irc{IRC\,+10216}

\bibpunct{(}{)}{;}{a}{}{,}

\begin{document}
  \title{Silicon in the dust formation zone of IRC\,+10216 as observed with PACS and SPIRE on board Herschel
\thanks{Herschel is an ESA space observatory with science instruments provided by European-led Principal Investigator consortia and with important participation from NASA. %It is open for proposals for observing time from the worldwide astronomical community.
}}

 \author{
         L. Decin\inst{1,2}
  	\and 
	J. Cernicharo\inst{3}
       \and
         M.J. Barlow\inst{4}
         \and
         P. Royer\inst{1}
          \and
         B. Vandenbussche\inst{1}
        \and
         R. Wesson\inst{4}
        \and
         E.T. Polehampton\inst{5,6}
        \and
	E. De Beck\inst{1}
	\and
	M. Ag\'undez\inst{3,10}
         \and
         J.A.D.L. Blommaert\inst{1}
         \and
         M. Cohen\inst{8}
	\and
	F. Daniel\inst{3}
         \and
         W. De Meester\inst{1}
         \and
         K. Exter\inst{1}
         \and
         H. Feuchtgruber\inst{11}
	\and
	J.P. Fonfr\'{\i}a\inst{7}
         \and
         W.K. Gear\inst{12}
 	\and
	J.R. Goicoechea\inst{3}
        \and
         H.L. Gomez\inst{12}
         \and
         M.A.T. Groenewegen\inst{13}
         \and
         P.C. Hargrave\inst{12}
         \and
         R. Huygen\inst{1}
         \and
         P. Imhof\inst{14}
         \and
         R.J. Ivison\inst{15}
         \and
         C. Jean\inst{1}
	\and
	F. Kerschbaum\inst{17}
         \and
         S.J. Leeks\inst{5}
         \and
         T. Lim\inst{5}
         \and
	M. Matsuura\inst{4,18}
	\and
         G. Olofsson\inst{16}
	\and
	T. Posch\inst{17}
         \and
         S. Regibo\inst{1}
         \and
         G. Savini\inst{4}
         \and
         B. Sibthorpe\inst{15}
         \and
         B.M. Swinyard\inst{5}
	\and	
	B. Tercero\inst{3}
          \and
         C. Waelkens\inst{1}
         \and
         D.K. Witherick\inst{4}
	\and
	J.A. Yates\inst{4}
         }

%  \offprints{}

  \institute{
	%1	
Instituut voor Sterrenkunde,
             Katholieke Universiteit Leuven, Celestijnenlaan 200D, 3001 Leuven, Belgium\\
	      \email{Leen.Decin@ster.kuleuven.be}\\
%2
 		\and
	Sterrenkundig Instituut Anton Pannekoek, University of Amsterdam, Science Park 904, NL-1098 Amsterdam, The Netherlands \\
%3
	\and
	   Laboratory of Molecular Astrophysics, Department of Astrophysics, CAB, INTA-CSIC, Ctra de Ajalvir, km 4, 28850 Torrejón de Ardoz, Madrid Spain \\
%4
       \and
             Dept of Physics \& Astronomy, University College London, Gower St, London WC1E 6BT, UK\\
%5
       \and
             Space Science and Technology Department, Rutherford Appleton Laboratory, Oxfordshire, OX11 0QX, UK\\
%6
       \and
            Department of Physics, University of Lethbridge, Lethbridge, Alberta, T1J 1B1, Canada\\
%7
	\and
Departamento de Astrofísica Molecular e Infrarroja, Instituto de Estructura de la Materia, CSIC, Serrano 121, 28006 Madrid, Spain\\
%8
      \and
             Radio Astronomy Laboratory, University of California at Berkeley, CA 94720, USA\\
%9
	\and
	Observatoire de Paris-Meudon, LERMA UMR CNRS 8112, 5 place Jules Janssen, 92195 Meudon Cedex, France \\
%10
	\and
	LUTH, Observatoire de Paris-Meudon, 5 Place Jules Janssen, 92190 Meudon, France \\
%11
	\and
             Max-Planck-Institut f\"ur extraterrestrische Physik, Giessenbachstrasse, 85748, Germany\\
%12
        \and
             School of Physics and Astronomy, Cardiff University, Queens Buildings, The Parade, Cardiff, CF24 3AA, UK\\
%13
	\and
	Royal Observatory of Belgium, Ringlaan 3, B-1180 Brussels, Belgium \\
%14
        \and
            Blue Sky Spectroscopy, 9/740 4 Ave S, Lethbridge, Alberta T1J 0N9, Canada \\
%15
	\and
	UK Astronomy Technology Centre, Royal Observatory Edinburgh, Blackford Hill, Edinburgh EH9 3HJ, UK \\
%16 
         \and
             Dept of Astronomy, Stockholm University, AlbaNova University Center, Roslagstullsbacken 21, 10691 Stockholm, Sweden\\
 %17 
         \and
    	    University of Vienna, Department of Astronomy, T{\"u}rkenschanzstra\ss{}e 17, A-1180 Vienna, Austria \\
%18
	\and
	Mullard Space Science Laboratory, University College London, 
Holmbury St. Mary, Dorking, Surrey RH5 6NT, UK
}

%  \offprints{}

  \date{Received / accepted }

%\abstract{}{}{}{}{}
% 5 {} token are mandatory
 \abstract{The interstellar medium is enriched primarily by matter ejected from evolved low and intermediate mass stars. The outflows from these stars create a circumstellar envelope in which a rich gas-phase and  dust-nucleation chemistry takes place. We observed the nearest carbon-rich evolved star, IRC\,+10216, using the PACS (55-210\,$\mu$m) and SPIRE (194-672\,$\mu$m) spectrometers  on board Herschel. We find several tens of lines from SiS and  SiO, including lines from the v=1 vibrational level. For SiS these transitions range up to J=124--123, corresponding to energies around 6700\,K, while the highest detectable transition is J=90--89 for SiO, which corresponds to an energy around 8400\,K. Both species trace  the dust formation zone of IRC\,+10216, and the broad energy ranges involved in their detected transitions permit us to derive the physical properties of the gas and the particular zone in which each species has been formed. This allows us to check the accuracy of chemical thermodynamical equilibrium models and the suggested depletion of SiS and SiO due to accretion onto dust grains.}
{}
{}
{}
{}

%{The interstellar medium is enriched primarily by matter ejected from evolved low and intermediate mass stars. The dust-driven outflows from these stars create a circumstellar envelope in which a rich gas-phase and  dust-nucleation chemistry takes place.}
%{By observing an evolved star in the infrared between  60 -- 670\,$\mu$m, we want to study the thermophysical and chemical structure of the circumstellar envelope around the central target.}
%{We have observed the nearest carbon-rich evolved star, IRC\,+10216, using the PACS and SPIRE spectrometers on board Herschel.}
%{We find several tens of lines from SiS and  SiO, including lines from the v=1 vibrational level. For SiS these transitions range up to J=124--123, corresponding to energies around 6700\,K. For SiO the highest detectable transition 
%is J=90--89, which corresponds to an energy around 8400\,K. Both species trace  the dust formation zone of IRC\,+10216 and the broad energy ranges involved in their detected transitions permit us to derive
%the physical properties of the gas and the particular zone in which each species has been formed. This allows us to check the accuracy
%of chemical thermodynamical equilibrium models and the suggested depletion of SiS and SiO due to accretion onto dust grains.}
%{}

  \maketitle
%
%________________________________________________________________

\section{Introduction}
%\object{IRC\,+10216} (\object{CW\,Leo}) was during the IRC all-sky infrared (IR) survey \citep{Neugebauer1969tmss.book.....N}. 
\object{IRC\,+10216} (\object{CW\,Leo}) is the brightest non-Solar System object in the sky at 5\,$\mu$m.  It is the nearest \citep[D$\sim$150\,pc,][]{Crosas1997ApJ...483..913C} carbon-rich evolved star, and it serves as an archetype for the study of mass loss on the asymptotic giant branch (AGB). The star is losing mass at a rate of $\sim$1--$3\times10^{-5}$\,\Msun/yr \citep{Crosas1997ApJ...483..913C, Schoier2000A&A...359..586S}, producing a dense, dusty circumstellar envelope (CSE). To date, more than 60 molecules have been detected in the CSE of IRC\,+10216 \citep[e.g.,][]{Cernicharo2000A&AS..142..181C, He2008ApJS..177..275H}. Thermodynamic equilibrium and non-equilibrium reactions, photochemical reactions, ion-molecule reactions, and the condensation of dust grains establish the abundance stratifications throughout the envelope. It is likely that IRC\,+10216 is  in an advanced evolutionary stage, marking the transition from an AGB to a planetary nebula \citep{Skinner1998MNRAS.300L..29S}. A detailed study of its infrared emission spectrum can yield unique information on the thermophysical and chemical structure of the outflow and on the history of mass loss during this important evolutionary phase.

\vspace*{-2ex}
\section{Observations and data reduction}

Thanks to its high infrared brightness, IRC +10216 is an ideal
target for observation with Herschel  \citep{Pilbratt2010}. PACS and SPIRE spectroscopic observations were obtained in the context of the Guaranteed Time Key Programme ``Mass-loss of Evolved StarS'' (Groenewegen et al., \emph{in prep.}).

The PACS instrument, its in-orbit performance and calibration,
and its scientific capabilities are described in \citet{Poglitsch2010}. The PACS spectroscopic observations of \irc\
consist of full SED scans between 52 and 210\,$\mu$m obtained in a 3$\times$1 raster, i.e.\ a pointing
on the central object, and two pointings 30\arcsec\ either side. The observations were performed on 2009 Nov 12 (OD\,182). The position angle was 110 degrees.
%The exact positions on the sky can be found in Decin et al. (2010, in prep).
The instrument mode was a non-standard version of the chop-nod PACS-SED AOT, used with a large chopper throw (6\,\arcmin). The spectral resolving power varies between 1000 and 4500. A description of the observing mode and of the data reduction process can be found in \citet{Royer2010}. The only difference with the data reduction of VY~CMa as presented in \citet{Royer2010} is that  the ground-based calibration was used for \irc. The estimated calibration uncertainty on the line fluxes is 50\,\%.

The SPIRE FTS measures the Fourier transform of the source spectrum across short (SSW, 194--313\,$\mu$m) and long (SLW, 303--671$\mu$m) wavelength bands simultaneously. The FWHM beamwidths of the SSW and SLW arrays vary between 17-19\arcsec and 29-42\arcsec respectively. The source spectrum, including the continuum, is restored by taking the inverse transform of the observed interferogram. The absolute flux calibration uncertainty is 15-20\% in the SSW band and 20-30\% in the SLW band above 20\,cm$^{-1}$ (up to 50\% below 20\,cm$^{-1}$). For more details on the SPIRE FTS and its calibration see \citet{Griffin2010} and \citet{Swinyard2010}.

IRC\,+10216 was observed with the high-resolution mode of the SPIRE FTS on the 2009 Nov 19  (OD 189). Twenty repetitions were used, each of  which consisted of one forward and one reverse scan of the FTS, with each scan taking 66.6s. The total on-source integration time was therefore 2664s. The unapodized spectral resolution is 1.4\,GHz (0.048\,cm$^{-1}$), and this is 2.1\,GHz (0.07\,cm$^{-1}$) after apodization \citep[using extended Norton-Beer function 1.5;][]{Naylor2007JOSAA..24.3644N}. 

PACS and SPIRE photometry observations of \irc\ are presented in \citet{Ladjal2010}.

\begin{figure*}
\begin{center}
 \includegraphics[height=0.625\textheight]{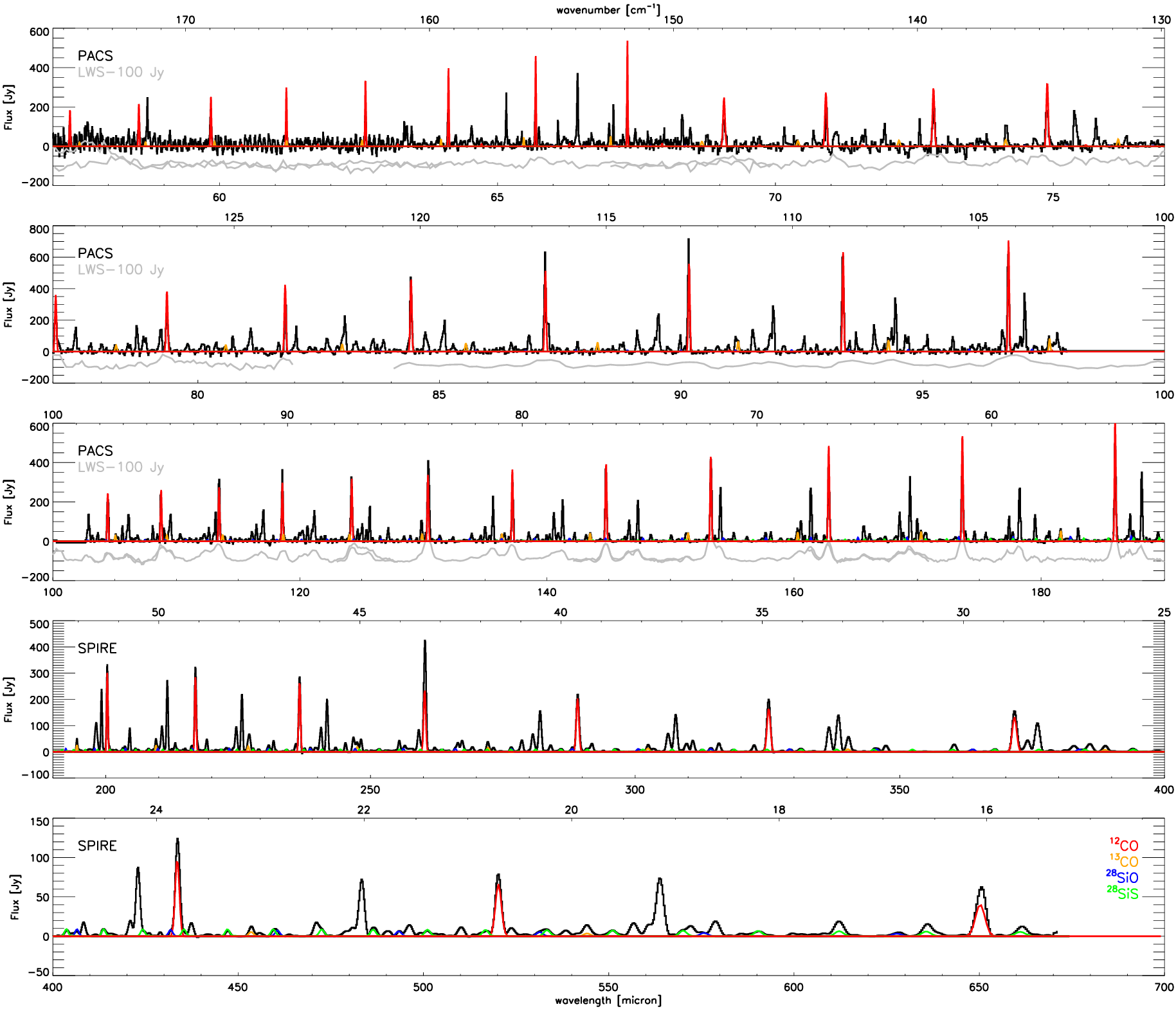}\\

\vspace*{3ex}
 \includegraphics[height=0.125\textheight]{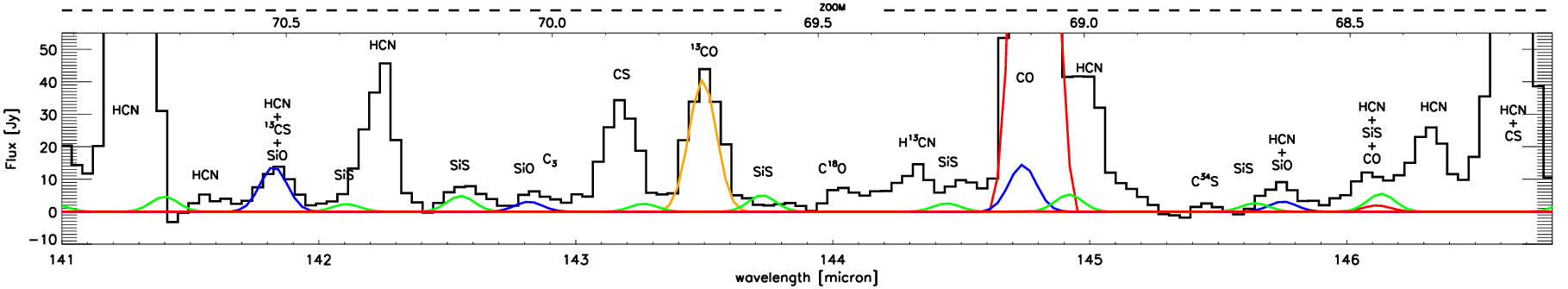}
\caption{Continuum-subtracted PACS and SPIRE spectrum of IRC\,+10216. In the three upper panels, the PACS spectrum of IRC\,+10216 (black) is compared to the ISO-LWS spectrum \citep[grey,][]{Cernicharo1996A&A...315L.201C}. The fourth and fifth panels show the SPIRE spectrum of IRC\,+10216 (black). The bottom panel zooms in on the 141 -- 146.8\,$\mu$m  region, where we identified the main molecular features. Theoretical line predictions  for $^{12}$CO (red), $^{13}$CO (orange), $^{28}$SiO (blue), and $^{28}$SiS (green) using the parameters as given in Table~\ref{Table:1} are displayed in all panels.}
\label{fig:1}
\end{center}
\end{figure*}

\section{Results} \label{results}

Currently, more than 500 molecular emission lines have been identified in the PACS and SPIRE spectra of \irc\ (see Fig.~\ref{fig:1}), belonging to 10 different molecules and their isotopologues ($^{12}$CO, $^{13}$CO, C$^{18}$O, H$^{12}$CN, H$^{13}$CN, H$_2$O, NH$_3$, SiS, SiO, CS, C$^{34}$S, $^{13}$CS, C$_3$, C$_2$H, HCl, and H$^{37}$Cl). The detection of this last molecule is discussed by \citet{Cernicharo2010}. In the ISO-LWS spectrum shown in Fig.~\ref{fig:1}, 57 lines belonging to CO and HCN were identified by \citet{Cernicharo1996A&A...315L.201C}. The number of identified lines increases to 280 in the PACS spectrum thanks to its higher spectral resolution. Most of the lines in the PACS and SPIRE spectrum arise from HCN, with the strongest lines from $^{12}$CO. HCN is one of the most abundant molecular species in the CSEs of carbon stars \citep{Willacy1998A&A...330..676W} and it is known to show maser action in various vibrational states.  The strength of the $^{12}$CO lines are diagnostics for the thermophysical structure (see Sect.~\ref{envelope_structure}). In this paper, we focus on the silicon-bearing molecules SiS and SiO, two refractory species that are formed in the inner envelope. As soon as the temperature of the gas falls below a certain critical value, the molecules can start to condense and form dust grains.

High-J rotational lines have been detected from both molecules. For SiO, 80 rotational transitions in the ground-state from J\,=\,11--10 to J\,=\,90--89 ($\rm{E_{up}\,=\,8432\,K}$), and 99 lines from J\,=\,26--25 to J\,=\,124--123 ($\rm{E_{up}\,=\,6678\,K}$) for SiS are clearly detected. From the detected lines, $\sim$45\% of both species is unblended (see Table~\ref{identifications} in the online Appendix, which also lists the detected $^{12}$CO and $^{13}$CO lines).
The emission lines of higher-J transitions and rotational transitions in the first vibrational state are very weak, but their line contribution can be deduced from the theoretical modelling (see Sect.~\ref{SiOandSiS}, and Table~\ref{identifications}). The line formation region of the highest-J lines of SiO (SiS)  is within the first 5\,\Rstar\ (10\,\Rstar), i.e., tracing the recently identified dust formation region \citep{Fonfria2008ApJ...673..445F}. %. The low J-lines in the SPIRE spectrum enable us to study the CSE until 1600\,\Rstar\ for SiO and until 1200\,\Rstar\ for SiS.

\subsection{Thermophysical structure of the envelope} \label{envelope_structure}
The large number of optically thick $^{12}$CO and optically thin $^{13}$CO lines enabled us to perform a tomographical study of the CSE. Properties of the circumstellar gas, such as the kinetic temperature, velocity, and density structure, were determined through a non-local thermodynamic equilibrium (non-LTE) radiative transfer modelling of the $^{12}$CO lines. The $^{12}$CO lines cover energy levels from J\,=\,3 (at 31\,K) to J\,=\,47 (at 5853\,K) and  trace the envelope for radii R$\,<\,1\,\times\,10^{17}$cm (R$<$2000\,\Rstar). The GASTRoNOoM code was used to calculate the kinetic temperature and velocity structure in the envelope and to solve the non-LTE radiative transfer equations \citep{Decin2006A&A...456..549D, Decin2010}. The rate equations were solved for the ground and first excited vibrational state, with J$^{\rm up}_{\rm max}$\,=\,60. The CO line list and collisional rates are discussed in \citet{Decin2010}.
The terminal velocity was deduced from ground-based observations of low-J $^{12}$CO lines \citep{Debeck}. The GASTRoNOoM code computes the velocity structure by solving the momentum equation and the temperature structure from the equation expressing the conservation of energy \citep[see Eq.~6 in][]{Decin2006A&A...456..549D}. However, the resulting temperature was slightly too low beyond 60\,\Rstar\ to correctly predict the lower excitation $^{12}$CO lines, which mainly reflects uncertainties in the gas-grain collisional heating. Therefore, we opted to use $T(R) \propto R^{-0.5}$ for $R>60$\,\Rstar. 

\begin{figure}[htp]
 \begin{center}
  \includegraphics[width=0.49\textwidth,angle=0]{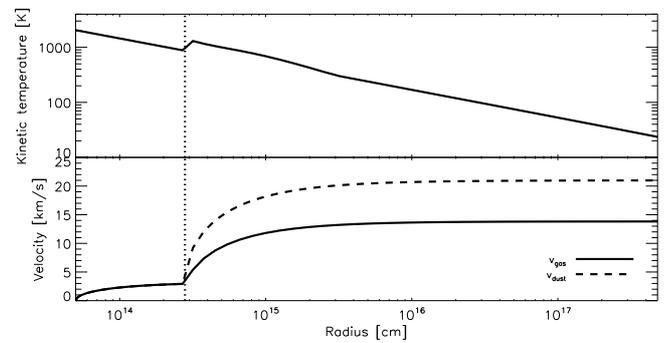}
\vspace*{1ex}
\caption{Thermodynamical structure in the envelope of IRC\,+10216 as derived from the $^{12}$CO rotational line transitions. The vertical dotted line represents the dust condensation radius. }
\label{fig:structure}
 \end{center}
\end{figure}

The best-fit model was determined using the log-likelihood function as described in \citet{Decin2007A&A...475..233D}.
The derived (circum)stellar parameters are given in Table~\ref{Table:1}, the deduced thermodynamical structure is displayed in Fig.~\ref{fig:structure}, and the line predictions are shown in Fig.~\ref{fig:1}. Specifically, we obtained a mass loss rate of $1 \times 10^{-5}$\,\Msun/yr (with an uncertainty of a factor 2) and a $^{12}$CO/$^{13}$CO ratio of $\sim$30$\pm5$.  The latter is on the lower side of the range of $^{12}$C/$^{13}$C ratios quoted in the literature, going from 20 \citep{Barnes1977ApJ...213...71B} to 50 \citep{Schoier2000A&A...359..586S}. The lowest value is obtained from vibra-rotational transitions in the fundamental band of CO, and higher values are often obtained from low-excitation CO or CS lines. The accuracy of isotopologue ratios obtained from low-excitation rotational transitions is often limited by the uncertain effect of photodissociation by interstellar UV photons and chemical fractionation \citep[e.g.,][]{Mamon1988ApJ...328..797M}, effects that are not hampering the high-excitation $^{12}$CO and $^{13}$CO lines in the PACS and SPIRE spectra.

\begin{table}
 \caption{Parameters for the best-fit model, where numbers in italics indicate input parameters that have been kept fixed at the given value.}
\label{Table:1}
\begin{center}
\vspace*{-1.5ex}
\begin{tabular}{lc|lc}
\hline \hline
\rule[0mm]{0mm}{5mm}\Teff [K] & \emph{2050}$^{a}$ & \Mdot [{\Msun}/yr] & $1 \times 10^{-5}$\\
$R_{\star}$ [$10^{13}$\,cm] & 5 & $R_{\rm{dust}}$ [\Rstar] & \emph{5.6}$^d$  \\
$[$CO/H$_2]$ [$10^{-3}$]    & \emph{1}$^{b}$ &  $^{12}$CO/$^{13}$CO & 30 \\
distance  [pc] & \emph{150}$^c$ &  n(SiO)/n(H$_2$) & $1 \times 10^{-7}$  \\ 
$v_{\infty}$     [km\,s$^{-1}$] &  \emph{14.5} &  n(SiS)/n(H$_2$) & $4 \times 10^{-6}$\\
\hline
\end{tabular}
\end{center}
$^{a}$\citet{Gonzalez2007ApJ...669..412G}, $^{b}$\citet{Zuckerman1986ApJ...304..394Z}, $^c$\citet{Crosas1997ApJ...483..913C}, $^d$\citet{Ridgway1988ApJ...326..843R}
\end{table}

\vspace*{-6ex}
\subsection{Abundance profiles of SiO and SiS} \label{SiOandSiS}
The SiO and SiS emission lines are modelled with the thermodynamical structure as deduced in Sect.~\ref{envelope_structure}. Linelists and (available) collisional rates are described in \citet{Decin2010}.  However, the lack of collisional rates for high-J transitions of both molecules with He or H$_2$ led us  calculate the level populations in LTE. This approach is justified since most of the detected high-J lines originate in the stellar photosphere and in the inner wind envelope, where the high gas density and temperature ensure thermal equilibrium for the level populations.  Pulsation driven shocks in the inner envelope may alter abundances predicted from equilibrium chemistry. The estimated uncertainty on the derived abundances  is a factor of 5, when taking the line flux uncertainty into account.

\begin{figure}[htp]
 \includegraphics[width=.48\textwidth]{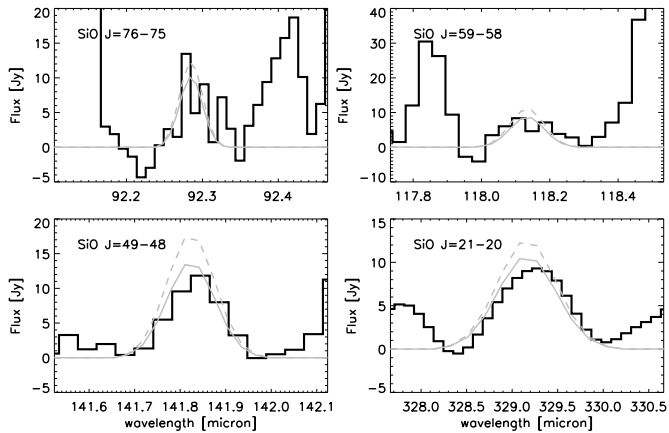}
\caption{Comparison between few PACS and SPIRE SiO v=0 lines (black) and theoretical line predictions (grey). Full grey lines represent theoretical line profiles using a constant SiO fractional abundance of [SiO/H$_2$]\,=\,$1\times10^{-7}$, and dashed grey lines represent model predictions simulating an inner wind abundance of [SiO/H$_2$]\,=\,$3\times10^{-7}$ for $R<8$\,\Rstar\ and $1\times10^{-7}$ beyond that radius.}
\label{fig:2}
\end{figure}

\paragraph{SiO:} 
%Spatial SiO J=2-1 maps  by \citet{Olofsson1982A&A...107..128O} indicate that the SiO emission region is $\la$12\arcsec\ in radius (or 560\,\Rstar). \citet{Lucas1992A&A...262..491L} derived a photodissocation radius of $\sim$7\arcsec (315\,\Rstar), while \citet{Schoier2006ApJ...649..965S} obtained $2.4\times10^{16}$cm (480\,\Rstar). 
Using an outer radius value of 560\,\Rstar\ \citep{Olofsson1982A&A...107..128O}, the derived fractional abundance is [SiO/H$_2$]\,=\,$1\times10^{-7}$, when assuming a constant abundance profile. The high-J SiO lines in the PACS and SPIRE spectrum provide us with a diagnostic tool for deducing  possible depletion from the gas from accretion onto dust grains. Unfortunately, the low signal-to-noise ratio of the (weak) high-excitation SiO lines prohibit us from putting strong constraints on the role of SiO in the dust formation around IRC\,+10216. 
When allowing for variations in the abundance profile \citep[as described in][]{Decin2010}, we deduce that the SiO fractional abundance in the inner wind  (R$\la$8\,\Rstar) can range between $0.2-3\times10^{-7}$, with the fractional abundance being $1\times10^{-7}$ beyond 8\,\Rstar\ (see Fig.~\ref{fig:2}).
\citet{Keady1993ApJ...406..199K}  derived an inner wind SiO abundance of $8\times10^{-7}$ from infrared ro-vibrational transitions. From low-excitation SiO lines, \citet{Schoier2006ApJ...649..965S} obtained an SiO  abundance in the region between $\sim$3 and 8\,\Rstar, as high as $\sim1.5\times10^{-6}$, superposed on a more spatially  extended region of 480\,\Rstar\ with a fractional abundance of $1.7\times10^{-7}$. The abundance in this compact inner-wind region is a factor 5 higher than our maximum deduced value of $3\times10^{-7}$ in the inner wind. The theoretically calculated photospheric TE value of SiO in carbon-rich envelopes is $\sim2.8\times10^{-8}$ \citep{Willacy1998A&A...330..676W}.  In their study of the effect of pulsationally induced non-chemical equilibrium in the inner wind of \irc, \citet{Willacy1998A&A...330..676W} obtained a fractional abundance of $3.8\times10^{-7}$. Since TE-value agrees with our minimum deduced value in the inner wind of $2 \times 10^{-8}$, and the non-TE value with the maximum deduced value, the effect of pulsationally induced non-equilibrium chemistry is difficult to estimate.

\vspace*{-1cm}
\paragraph{SiS:}  The SiS fractional abundances derived from the PACS and SPIRE observations is [SiS/H$_2$]\,=\,$4\times10^{-6}$. When taking the abundance uncertainty into account, this agrees with the results of \citet{Schoier2007A&A...473..871S}, who find a value of $2\times10^{-6}$ from low-excitation SiS lines. From observations of the 13.5\,$\mu$m fundamental band of SiS, \citet{Boyle1994ApJ...420..863B} obtained a gradient in the abundance of SiS, going from $4.3\times10^{-6}$ at a distance of 12\,\Rstar\ and rising to $4.3\times10^{-5}$ close to the stellar surface. \citet{Bieging1989ApJ...343L..25B} obtained a much lower value of [SiS/H$_2$]\,=\,$7.5\times10^{-6}$ for $R<3\times10^{15}$\,cm and $6.5\times10^{-7}$ beyond that radius. The PACS and SPIRE observations can allow  for a change of a factor of $2$ in the first few stellar radii (i.e, minimum of $2 \times 10^{-6}$, maximum of $8 \times 10^{-6}$ for $R\la12$\,\Rstar). The SiS TE and non-TE inner wind values as computed by \citet{Willacy1998A&A...330..676W} are [SiS/H$_2$]\,=\,$1.5\times10^{-5}$ and $3.4\times10^{-5}$, respectively. That our deduced SiS fractional abundance is clearly lower than the non-TE value of \citet{Willacy1998A&A...330..676W} might indicate toward uncertainties in the estimated rate values in Eqs.~(11)-(13) in \citet{Willacy1998A&A...330..676W}.

\vspace*{-1.5ex}
\section{Conclusion}
The PACS and SPIRE spectroscopic observations of  \irc\ have been shown to be of excellent quality for studying the thermodynamical and chemical structure of the envelope, created by its copious mass loss. The temperature and mass-loss rate of the envelope are derived from the $^{12}$CO lines. Both SiO and SiS are refractory species, and the PACS and SPIRE data can provide a strong diagnostic tool for determining their role in the dust formation process. 
Analysing the high-J SiO and SiS lines yields a constant fractional abundance of $1\times10^{-7}$ and $4\times10^{-6}$, respectively. 
However, we detect only v=0 and v=1 transitions for both species, mainly because of the densely populated spectrum of IRC+10216, while it is known from ground-based observations that levels of SiS up to v=8 have been detected (Ag\'undez et al., 2010, \emph{in prep.}). Moreover, the low-J transitions of SiO and SiS, which are more sensitive to the external envelope, are not accesible to PACS and SPIRE. Since the high-J lines in the ground-state and the v=1 lines of both molecules are very weak, we cannot put strong constraints on the fractional abundance in the inner envelope (R$\la10$\,\Rstar). For SiO, 1/3 at most is estimated to take part in dust formation process, while we deduce a fraction of 1/2 for SiS.
Only a merged set of millimeter, submillimeter, and far-infrared observations of SiO and SiS can provide a detailed analysis of the abundance of these species from the photosphere to the photodissociation zone (Ag\'undez et al., 2010, \emph{in prep.}).

\vspace*{-1ex}
\begin{acknowledgements}
PACS was developed by a consortium of institutes led by MPE (Germany) and including UVIE (Austria); KUL, CSL, IMEC (Belgium); CEA, OAMP (France); MPIA (Germany); IFSI, OAP/AOT, OAA/CAISMI, LENS, SISSA (Italy); IAC (Spain). This development has been supported by the funding agencies BMVIT (Austria), ESA-PRODEX (Belgium), CEA/CNES (France), DLR (Germany), ASI (Italy), and CICT/MCT (Spain). SPIRE has been developed by a consortium of institutes led by
Cardiff Univ. (UK) and including Univ. Lethbridge (Canada);
NAOC (China); CEA, LAM (France); IFSI, Univ. Padua (Italy);
IAC (Spain); Stockholm Observatory (Sweden); Imperial College
London, RAL, UCL-MSSL, UKATC, Univ. Sussex (UK); Caltech, JPL,
NHSC, Univ. Colorado (USA). This development has been supported
by national funding agencies: CSA (Canada); NAOC (China); CEA,
CNES, CNRS (France); ASI (Italy); MCINN (Spain); SNSB (Sweden);
STFC (UK); and NASA (USA). LD  acknowledges financial support from the Fund for Scientific
  Research - Flanders (FWO). MG, DL, JB, WDM, KE, RH, CH, SR, PR, and BV acknowledge support from the Belgian Federal Science Policy Office via the PRODEX Programme of ESA. FK acknowledges funding by the Austrian Science Fund FWF under project number P18939-N16 and I163-N16

\end{acknowledgements}
\vspace*{-1ex}
%%%%%%%%%%%%%%%%%%%%%%%%%%%%%%%%%%%%%%%%%%%%%%%%%%%%%%%%%%%%%%%%%%%%%%%%%%%
\vspace*{-.5cm}
\bibliographystyle{aa}
\bibliography{14562}
%%%%%%%%%%%%%%%%%%%%%%%%%%%%%%%%%%%%%%%%%%%%%%%%%%%%%%%%%%%%%%%%%%%%%%%%%%%

\Online
\begin{appendix}
\section{Identified lines of $^{12}$C$^{16}$O, $^{13}$C$^{16}$O, $^{28}$Si$^{16}$O, and $^{28}$Si$^{32}$S}
% if table 2
\longtab{2}{
\begin{longtable}{lllllc}
\caption{\label{identifications} Identified lines of $^{12}$C$^{16}$O, $^{13}$C$^{16}$O, $^{28}$Si$^{16}$O, and $^{28}$Si$^{32}$S in the PACS and SPIRE spectrum of IRC\,10216. The integrated flux values in the fifth column are for the model predictions as shown in Fig.~\ref{fig:1}. The last column indicates whether a line is blended strongly with another line.}\\
\hline\hline
Molecule & Transition & Frequency & Wavelength & Integrated Flux & Blend \\
 &   & [GHz]  &[$\mu$m] & [erg/cm$^2$/s] & \\ 
\hline
\endfirsthead
\caption{continued.}\\
\hline\hline
Molecule & Transition & Frequency  & Wavelength & Integrated Flux & Blend \\
 &   & [GHz] &[$\mu$m] & [erg/cm$^2$/s] & \\ 
\hline
\endhead
\hline
\endfoot
\hline
$^{12}$C$^{16}$O & v=0-0, J=4-3 &  461.042 &  650.250 & 1.35e-12 & no \\
$^{12}$C$^{16}$O & v=0-0, J=5-4 &  576.267 &  520.232 & 2.20e-12 & no \\
$^{12}$C$^{16}$O & v=0-0, J=6-5 &  691.474 &  433.555 & 3.08e-12 & no \\
$^{12}$C$^{16}$O & v=0-0, J=7-6 &  806.652 &  371.650 & 4.17e-12 & no \\
$^{12}$C$^{16}$O & v=0-0, J=8-7 &  921.799 &  325.225 & 4.98e-12 & no \\
$^{12}$C$^{16}$O & v=0-0, J=9-8 & 1036.913 &  289.120 & 6.01e-12 & no \\
$^{12}$C$^{16}$O & v=0-0, J=10-9 & 1151.986 &  260.240 & 6.67e-12 & no \\
$^{12}$C$^{16}$O & v=0-0, J=11-10 & 1267.016 &  236.613 & 7.22e-12 & no \\
$^{12}$C$^{16}$O & v=0-0, J=12-11 & 1381.995 &  216.927 & 1.98e-12 & no \\
$^{12}$C$^{16}$O & v=0-0, J=13-12 & 1496.924 &  200.272 & 2.99e-12 & no \\
$^{12}$C$^{16}$O & v=0-0, J=14-13 & 1611.792 &  185.999 & 7.36e-12 & no \\
$^{12}$C$^{16}$O & v=0-0, J=15-14 & 1726.604 &  173.631 & 7.31e-12 & no \\
$^{12}$C$^{16}$O & v=0-0, J=16-15 & 1841.347 &  162.812 & 7.32e-12 & no \\
$^{12}$C$^{16}$O & v=0-0, J=17-16 & 1956.017 &  153.267 & 6.95e-12 & no \\
$^{12}$C$^{16}$O & v=0-0, J=18-17 & 2070.616 &  144.784 & 6.68e-12 & no \\
$^{12}$C$^{16}$O & v=0-0, J=19-18 & 2185.134 &  137.196 & 6.47e-12 & no \\
$^{12}$C$^{16}$O & v=0-0, J=20-19 & 2299.570 &  130.369 & 6.15e-12 & no \\
$^{12}$C$^{16}$O & v=0-0, J=21-20 & 2413.917 &  124.193 & 5.88e-12 & no \\
$^{12}$C$^{16}$O & v=0-0, J=22-21 & 2528.171 &  118.581 & 5.53e-12 & no \\
$^{12}$C$^{16}$O & v=0-0, J=23-22 & 2642.332 &  113.458 & 5.08e-12 & no \\
$^{12}$C$^{16}$O & v=0-0, J=24-23 & 2756.388 &  108.763 & 4.83e-12 & no \\
$^{12}$C$^{16}$O & v=0-0, J=25-24 & 2870.339 &  104.445 & 4.51e-12 & no \\
$^{12}$C$^{16}$O & v=0-0, J=27-26 & 3097.909 &   96.773 & 4.06e-12 & no \\
$^{12}$C$^{16}$O & v=0-0, J=28-27 & 3211.518 &   93.349 & 3.85e-12 & no \\
$^{12}$C$^{16}$O & v=0-0, J=29-28 & 3325.005 &   90.163 & 3.55e-12 & yes \\
$^{12}$C$^{16}$O & v=0-0, J=30-29 & 3438.365 &   87.190 & 3.38e-12 & yes \\
$^{12}$C$^{16}$O & v=0-0, J=31-30 & 3551.594 &   84.411 & 3.09e-12 & no \\
$^{12}$C$^{16}$O & v=0-0, J=32-31 & 3664.685 &   81.806 & 2.95e-12 & no \\
$^{12}$C$^{16}$O & v=0-0, J=33-32 & 3777.637 &   79.360 & 2.68e-12 & no \\
$^{12}$C$^{16}$O & v=0-0, J=34-33 & 3890.443 &   77.059 & 2.59e-12 & no \\
$^{12}$C$^{16}$O & v=0-0, J=35-34 & 4003.102 &   74.890 & 2.31e-12 & no \\
$^{12}$C$^{16}$O & v=0-0, J=36-35 & 4115.605 &   72.843 & 2.18e-12 & no \\
$^{12}$C$^{16}$O & v=0-0, J=37-36 & 4227.953 &   70.907 & 2.05e-12 & yes \\
$^{12}$C$^{16}$O & v=0-0, J=38-37 & 4340.138 &   69.074 & 1.89e-12 & no \\
$^{12}$C$^{16}$O & v=0-0, J=39-38 & 4452.159 &   67.336 & 1.74e-12 & no \\
$^{12}$C$^{16}$O & v=0-0, J=40-39 & 4564.005 &   65.686 & 1.60e-12 & no \\
$^{12}$C$^{16}$O & v=0-0, J=41-40 & 4675.681 &   64.117 & 1.50e-12 & no \\
$^{12}$C$^{16}$O & v=0-0, J=42-41 & 4787.174 &   62.624 & 1.34e-12 & no \\
$^{12}$C$^{16}$O & v=0-0, J=43-42 & 4898.484 &   61.201 & 1.31e-12 & no \\
$^{12}$C$^{16}$O & v=0-0, J=44-43 & 5009.608 &   59.843 & 1.18e-12 & yes \\
$^{12}$C$^{16}$O & v=0-0, J=45-44 & 5120.540 &   58.547 & 1.12e-12 & no \\
\hline
$^{12}$C$^{16}$O & v=1-1, J=28-27 & 3182.136 &   94.211 & 1.42e-13 & yes \\
$^{12}$C$^{16}$O & v=1-1, J=29-28 & 3294.573 &   90.996 & 1.41e-13 & yes \\
$^{12}$C$^{16}$O & v=1-1, J=30-29 & 3406.884 &   87.996 & 1.51e-13 & no \\
$^{12}$C$^{16}$O & v=1-1, J=31-30 & 3519.060 &   85.191 & 1.48e-13 & no \\
$^{12}$C$^{16}$O & v=1-1, J=32-31 & 3631.105 &   82.562 & 1.51e-13 & no \\
\hline
$^{13}$C$^{16}$O & v=0-0, J=5-4 &  550.926 &  544.161 & 1.17e-13 & yes \\
$^{13}$C$^{16}$O & v=0-0, J=6-5 &  661.066 &  453.498 & 1.75e-13 & no \\
$^{13}$C$^{16}$O & v=0-0, J=7-6 &  771.183 &  388.743 & 2.39e-13 & yes \\
$^{13}$C$^{16}$O & v=0-0, J=8-7 &  881.273 &  340.181 & 3.19e-13 & yes \\
$^{13}$C$^{16}$O & v=0-0, J=9-8 &  991.330 &  302.414 & 4.05e-13 & no \\
$^{13}$C$^{16}$O & v=0-0, J=10-9 & 1101.351 &  272.204 & 4.85e-13 & no \\
$^{13}$C$^{16}$O & v=0-0, J=11-10 & 1211.330 &  247.490 & 5.39e-13 & no \\
$^{13}$C$^{16}$O & v=0-0, J=12-11 & 1321.267 &  226.898 & 6.11e-13 & yes \\
$^{13}$C$^{16}$O & v=0-0, J=13-12 & 1431.152 &  209.476 & 6.27e-13 & no \\
$^{13}$C$^{16}$O & v=0-0, J=14-13 & 1540.987 &  194.546 & 2.91e-13 & no \\
$^{13}$C$^{16}$O & v=0-0, J=15-14 & 1650.768 &  181.608 & 6.94e-13 & no \\
$^{13}$C$^{16}$O & v=0-0, J=16-15 & 1760.487 &  170.290 & 6.99e-13 & no \\
$^{13}$C$^{16}$O & v=0-0, J=17-16 & 1870.142 &  160.305 & 7.05e-13 & no \\
$^{13}$C$^{16}$O & v=0-0, J=18-17 & 1979.728 &  151.431 & 7.03e-13 & no \\
$^{13}$C$^{16}$O & v=0-0, J=19-18 & 2089.239 &  143.494 & 6.87e-13 & no \\
$^{13}$C$^{16}$O & v=0-0, J=20-19 & 2198.678 &  136.351 & 6.68e-13 & no \\
$^{13}$C$^{16}$O & v=0-0, J=21-20 & 2308.034 &  129.891 & 6.32e-13 & yes \\
$^{13}$C$^{16}$O & v=0-0, J=22-21 & 2417.308 &  124.019 & 6.36e-13 & yes \\
$^{13}$C$^{16}$O & v=0-0, J=23-22 & 2526.492 &  118.660 & 5.93e-13 & yes \\
$^{13}$C$^{16}$O & v=0-0, J=24-23 & 2635.584 &  113.748 & 5.73e-13 & no \\
$^{13}$C$^{16}$O & v=0-0, J=25-24 & 2744.579 &  109.231 & 5.63e-13 & no \\
$^{13}$C$^{16}$O & v=0-0, J=26-25 & 2853.476 &  105.062 & 5.39e-13 & no \\
$^{13}$C$^{16}$O & v=0-0, J=28-27 & 3070.949 &   97.622 & 4.90e-13 & no \\
$^{13}$C$^{16}$O & v=0-0, J=29-28 & 3179.518 &   94.289 & 4.75e-13 & no \\
$^{13}$C$^{16}$O & v=0-0, J=30-29 & 3287.974 &   91.178 & 4.58e-13 & no \\
$^{13}$C$^{16}$O & v=0-0, J=31-30 & 3396.307 &   88.270 & 4.39e-13 & no \\
$^{13}$C$^{16}$O & v=0-0, J=32-31 & 3504.517 &   85.545 & 4.22e-13 & no \\
$^{13}$C$^{16}$O & v=0-0, J=33-32 & 3612.599 &   82.985 & 4.11e-13 & yes \\
$^{13}$C$^{16}$O & v=0-0, J=34-33 & 3720.548 &   80.578 & 3.97e-13 & no \\
$^{13}$C$^{16}$O & v=0-0, J=35-34 & 3828.359 &   78.308 & 3.83e-13 & no \\
$^{13}$C$^{16}$O & v=0-0, J=36-35 & 3936.033 &   76.166 & 3.72e-13 & no \\
$^{13}$C$^{16}$O & v=0-0, J=37-36 & 4043.562 &   74.141 & 3.63e-13 & yes \\
$^{13}$C$^{16}$O & v=0-0, J=38-37 & 4150.945 &   72.223 & 3.54e-13 & yes \\
$^{13}$C$^{16}$O & v=0-0, J=39-38 & 4258.172 &   70.404 & 3.52e-13 & no \\
$^{13}$C$^{16}$O & v=0-0, J=40-39 & 4365.246 &   68.677 & 3.35e-13 & no \\
$^{13}$C$^{16}$O & v=0-0, J=41-40 & 4472.161 &   67.035 & 3.33e-13 & no \\
$^{13}$C$^{16}$O & v=0-0, J=42-41 & 4578.908 &   65.472 & 3.11e-13 & no \\
$^{13}$C$^{16}$O & v=0-0, J=43-42 & 4685.490 &   63.983 & 3.16e-13 & no \\
$^{13}$C$^{16}$O & v=0-0, J=44-43 & 4791.901 &   62.562 & 3.03e-13 & no \\
\hline
$^{28}$Si$^{16}$O & v=0-0, J=11-10 &  477.569 &  627.746 & 7.26e-14 & no \\
$^{28}$Si$^{16}$O & v=0-0, J=12-11 &  520.740 &  575.705 & 9.52e-14 & no \\
$^{28}$Si$^{16}$O & v=0-0, J=13-12 &  564.209 &  531.350 & 1.19e-13 & no \\
$^{28}$Si$^{16}$O & v=0-0, J=14-13 &  607.679 &  493.340 & 1.41e-13 & no \\
$^{28}$Si$^{16}$O & v=0-0, J=15-14 &  650.850 &  460.617 & 1.62e-13 & no \\
$^{28}$Si$^{16}$O & v=0-0, J=16-15 &  694.319 &  431.779 & 1.81e-13 & no \\
$^{28}$Si$^{16}$O & v=0-0, J=17-16 &  737.490 &  406.504 & 1.94e-13 & no \\
$^{28}$Si$^{16}$O & v=0-0, J=18-17 &  780.959 &  383.877 & 2.11e-13 & no \\
$^{28}$Si$^{16}$O & v=0-0, J=19-18 &  824.130 &  363.769 & 2.16e-13 & no \\
$^{28}$Si$^{16}$O & v=0-0, J=20-19 &  867.600 &  345.542 & 2.29e-13 & no \\
$^{28}$Si$^{16}$O & v=0-0, J=21-20 &  910.770 &  329.164 & 2.33e-13 & no \\
$^{28}$Si$^{16}$O & v=0-0, J=22-21 &  953.940 &  314.268 & 2.46e-13 & no \\
$^{28}$Si$^{16}$O & v=0-0, J=23-22 &  997.410 &  300.571 & 2.49e-13 & no \\
$^{28}$Si$^{16}$O & v=0-0, J=24-23 & 1040.580 &  288.101 & 2.57e-13 & no \\
$^{28}$Si$^{16}$O & v=0-0, J=25-24 & 1083.750 &  276.625 & 2.57e-13 & no \\
$^{28}$Si$^{16}$O & v=0-0, J=26-25 & 1126.920 &  266.028 & 2.55e-13 & no \\
$^{28}$Si$^{16}$O & v=0-0, J=27-26 & 1170.090 &  256.213 & 2.64e-13 & no \\
$^{28}$Si$^{16}$O & v=0-0, J=28-27 & 1213.260 &  247.097 & 2.66e-13 & yes \\
$^{28}$Si$^{16}$O & v=0-0, J=29-28 & 1256.430 &  238.606 & 2.60e-13 & no \\
$^{28}$Si$^{16}$O & v=0-0, J=30-29 & 1299.601 &  230.680 & 2.66e-13 & yes \\
$^{28}$Si$^{16}$O & v=0-0, J=31-30 & 1342.471 &  223.314 & 2.62e-13 & no \\
$^{28}$Si$^{16}$O & v=0-0, J=32-31 & 1385.641 &  216.357 & 7.17e-14 & yes \\
$^{28}$Si$^{16}$O & v=0-0, J=33-32 & 1428.811 &  209.820 & 2.73e-13 & no \\
$^{28}$Si$^{16}$O & v=0-0, J=34-33 & 1471.681 &  203.707 & 9.86e-14 & no \\
$^{28}$Si$^{16}$O & v=0-0, J=35-34 & 1514.852 &  197.902 & 2.75e-13 & yes \\
$^{28}$Si$^{16}$O & v=0-0, J=36-35 & 1557.722 &  192.456 & 1.22e-13 & no \\
$^{28}$Si$^{16}$O & v=0-0, J=37-36 & 1600.592 &  187.301 & 2.75e-13 & yes \\
$^{28}$Si$^{16}$O & v=0-0, J=38-37 & 1643.762 &  182.382 & 2.70e-13 & no \\
$^{28}$Si$^{16}$O & v=0-0, J=39-38 & 1686.633 &  177.746 & 2.73e-13 & yes \\
$^{28}$Si$^{16}$O & v=0-0, J=40-39 & 1729.503 &  173.340 & 2.77e-13 & no \\
$^{28}$Si$^{16}$O & v=0-0, J=41-40 & 1772.373 &  169.147 & 2.75e-13 & no \\
$^{28}$Si$^{16}$O & v=0-0, J=42-41 & 1814.944 &  165.180 & 2.79e-13 & no \\
$^{28}$Si$^{16}$O & v=0-0, J=43-42 & 1857.814 &  161.368 & 2.78e-13 & yes \\
$^{28}$Si$^{16}$O & v=0-0, J=44-43 & 1900.684 &  157.729 & 2.77e-13 & yes \\
$^{28}$Si$^{16}$O & v=0-0, J=45-44 & 1943.255 &  154.273 & 2.69e-13 & no \\
$^{28}$Si$^{16}$O & v=0-0, J=46-45 & 1986.125 &  150.943 & 2.68e-13 & no \\
$^{28}$Si$^{16}$O & v=0-0, J=47-46 & 2028.696 &  147.776 & 2.70e-13 & no \\
$^{28}$Si$^{16}$O & v=0-0, J=48-47 & 2071.266 &  144.739 & 2.73e-13 & yes \\
$^{28}$Si$^{16}$O & v=0-0, J=49-48 & 2113.837 &  141.824 & 2.64e-13 & no \\
$^{28}$Si$^{16}$O & v=0-0, J=50-49 & 2156.408 &  139.024 & 2.69e-13 & no \\
$^{28}$Si$^{16}$O & v=0-0, J=51-50 & 2198.978 &  136.333 & 2.60e-13 & yes \\
$^{28}$Si$^{16}$O & v=0-0, J=52-51 & 2241.549 &  133.743 & 2.62e-13 & no \\
$^{28}$Si$^{16}$O & v=0-0, J=53-52 & 2283.819 &  131.268 & 2.59e-13 & yes \\
$^{28}$Si$^{16}$O & v=0-0, J=54-53 & 2326.390 &  128.866 & 2.48e-13 & yes \\
$^{28}$Si$^{16}$O & v=0-0, J=55-54 & 2368.661 &  126.566 & 2.51e-13 & yes \\
$^{28}$Si$^{16}$O & v=0-0, J=56-55 & 2410.931 &  124.347 & 2.48e-13 & yes \\
$^{28}$Si$^{16}$O & v=0-0, J=57-56 & 2453.202 &  122.205 & 2.43e-13 & no \\
$^{28}$Si$^{16}$O & v=0-0, J=58-57 & 2495.473 &  120.135 & 2.37e-13 & yes \\
$^{28}$Si$^{16}$O & v=0-0, J=59-58 & 2537.744 &  118.133 & 2.27e-13 & no \\
$^{28}$Si$^{16}$O & v=0-0, J=60-59 & 2580.014 &  116.198 & 2.27e-13 & yes \\
$^{28}$Si$^{16}$O & v=0-0, J=61-60 & 2621.985 &  114.338 & 2.23e-13 & yes \\
$^{28}$Si$^{16}$O & v=0-0, J=62-61 & 2664.256 &  112.524 & 2.17e-13 & yes \\
$^{28}$Si$^{16}$O & v=0-0, J=63-62 & 2706.227 &  110.779 & 2.10e-13 & no \\
$^{28}$Si$^{16}$O & v=0-0, J=64-63 & 2748.198 &  109.087 & 2.05e-13 & yes \\
$^{28}$Si$^{16}$O & v=0-0, J=65-64 & 2790.169 &  107.446 & 2.01e-13 & no \\
$^{28}$Si$^{16}$O & v=0-0, J=66-65 & 2832.140 &  105.854 & 1.94e-13 & yes \\
$^{28}$Si$^{16}$O & v=0-0, J=67-66 & 2874.111 &  104.308 & 1.84e-13 & yes \\
$^{28}$Si$^{16}$O & v=0-0, J=68-67 & 2915.782 &  102.817 & 1.80e-13 & yes \\
$^{28}$Si$^{16}$O & v=0-0, J=72-71 & 3082.467 &   97.257 & 1.56e-13 & no \\
$^{28}$Si$^{16}$O & v=0-0, J=73-72 & 3124.138 &   95.960 & 1.52e-13 & yes \\
$^{28}$Si$^{16}$O & v=0-0, J=74-73 & 3165.809 &   94.697 & 1.46e-13 & yes \\
$^{28}$Si$^{16}$O & v=0-0, J=75-74 & 3207.180 &   93.475 & 1.37e-13 & yes \\
$^{28}$Si$^{16}$O & v=0-0, J=76-75 & 3248.552 &   92.285 & 1.34e-13 & no \\
$^{28}$Si$^{16}$O & v=0-0, J=77-76 & 3289.923 &   91.124 & 1.28e-13 & yes \\
$^{28}$Si$^{16}$O & v=0-0, J=78-77 & 3331.294 &   89.993 & 1.19e-13 & yes \\
$^{28}$Si$^{16}$O & v=0-0, J=79-78 & 3372.666 &   88.889 & 1.17e-13 & no \\
$^{28}$Si$^{16}$O & v=0-0, J=80-79 & 3413.737 &   87.819 & 1.13e-13 & yes \\
$^{28}$Si$^{16}$O & v=0-0, J=81-80 & 3454.809 &   86.775 & 1.08e-13 & yes \\
$^{28}$Si$^{16}$O & v=0-0, J=82-81 & 3496.180 &   85.749 & 1.03e-13 & yes \\
$^{28}$Si$^{16}$O & v=0-0, J=83-82 & 3536.952 &   84.760 & 9.76e-14 & yes \\
$^{28}$Si$^{16}$O & v=0-0, J=84-83 & 3578.024 &   83.787 & 9.20e-14 & yes \\
$^{28}$Si$^{16}$O & v=0-0, J=85-84 & 3619.095 &   82.836 & 8.84e-14 & yes \\
$^{28}$Si$^{16}$O & v=0-0, J=86-85 & 3659.867 &   81.913 & 8.23e-14 & yes \\
$^{28}$Si$^{16}$O & v=0-0, J=87-86 & 3700.639 &   81.011 & 8.11e-14 & yes \\
$^{28}$Si$^{16}$O & v=0-0, J=88-87 & 3741.410 &   80.128 & 7.74e-14 & yes \\
$^{28}$Si$^{16}$O & v=0-0, J=89-88 & 3782.182 &   79.264 & 7.36e-14 & yes \\
$^{28}$Si$^{16}$O & v=0-0, J=90-89 & 3822.654 &   78.425 & 7.13e-14 & yes \\
\hline
$^{28}$Si$^{16}$O & v=1-1, J=39-38 & 1674.641 &  179.019 & 4.06e-14 & yes \\
$^{28}$Si$^{16}$O & v=1-1, J=40-39 & 1717.211 &  174.581 & 4.28e-14 & yes \\
$^{28}$Si$^{16}$O & v=1-1, J=41-40 & 1759.782 &  170.358 & 4.40e-14 & yes \\
$^{28}$Si$^{16}$O & v=1-1, J=42-41 & 1802.353 &  166.334 & 4.74e-14 & no \\
$^{28}$Si$^{16}$O & v=1-1, J=43-42 & 1844.923 &  162.496 & 4.88e-14 & yes \\
$^{28}$Si$^{16}$O & v=1-1, J=44-43 & 1887.194 &  158.856 & 5.15e-14 & yes \\
$^{28}$Si$^{16}$O & v=1-1, J=45-44 & 1929.764 &  155.352 & 5.28e-14 & no \\
$^{28}$Si$^{16}$O & v=1-1, J=46-45 & 1972.035 &  152.022 & 5.48e-14 & yes \\
$^{28}$Si$^{16}$O & v=1-1, J=47-46 & 2014.606 &  148.809 & 5.53e-14 & yes \\
$^{28}$Si$^{16}$O & v=1-1, J=48-47 & 2056.876 &  145.751 & 5.64e-14 & yes \\
$^{28}$Si$^{16}$O & v=1-1, J=49-48 & 2099.147 &  142.816 & 5.82e-14 & no \\
\hline
$^{28}$Si$^{32}$S & v=0-0, J=26-25 &  471.608 &  635.682 & 1.29e-13 & yes \\
$^{28}$Si$^{32}$S & v=0-0, J=27-26 &  489.713 &  612.180 & 1.45e-13 & yes \\
$^{28}$Si$^{32}$S & v=0-0, J=28-27 &  507.813 &  590.360 & 1.47e-13 & no \\
$^{28}$Si$^{32}$S & v=0-0, J=29-28 &  525.910 &  570.045 & 1.61e-13 & no \\
$^{28}$Si$^{32}$S & v=0-0, J=30-29 &  544.003 &  551.087 & 1.61e-13 & no \\
$^{28}$Si$^{32}$S & v=0-0, J=31-30 &  562.091 &  533.352 & 1.76e-13 & no \\
$^{28}$Si$^{32}$S & v=0-0, J=32-31 &  580.175 &  516.728 & 1.74e-13 & no \\
$^{28}$Si$^{32}$S & v=0-0, J=33-32 &  598.254 &  501.112 & 1.85e-13 & no \\
$^{28}$Si$^{32}$S & v=0-0, J=34-33 &  616.328 &  486.417 & 1.84e-13 & no \\
$^{28}$Si$^{32}$S & v=0-0, J=35-34 &  634.398 &  472.562 & 1.91e-13 & no \\
$^{28}$Si$^{32}$S & v=0-0, J=36-35 &  652.463 &  459.478 & 1.93e-13 & no \\
$^{28}$Si$^{32}$S & v=0-0, J=37-36 &  670.522 &  447.103 & 1.93e-13 & no \\
$^{28}$Si$^{32}$S & v=0-0, J=38-37 &  688.576 &  435.380 & 1.98e-13 & no \\
$^{28}$Si$^{32}$S & v=0-0, J=39-38 &  706.625 &  424.260 & 1.93e-13 & yes \\
$^{28}$Si$^{32}$S & v=0-0, J=40-39 &  724.668 &  413.696 & 2.00e-13 & no \\
$^{28}$Si$^{32}$S & v=0-0, J=41-40 &  742.705 &  403.649 & 2.00e-13 & no \\
$^{28}$Si$^{32}$S & v=0-0, J=42-41 &  760.737 &  394.082 & 1.97e-13 & no \\
$^{28}$Si$^{32}$S & v=0-0, J=43-42 &  778.762 &  384.960 & 2.03e-13 & no \\
$^{28}$Si$^{32}$S & v=0-0, J=44-43 &  796.782 &  376.254 & 2.00e-13 & yes \\
$^{28}$Si$^{32}$S & v=0-0, J=45-44 &  814.795 &  367.936 & 1.99e-13 & no \\
$^{28}$Si$^{32}$S & v=0-0, J=46-45 &  832.801 &  359.981 & 2.04e-13 & yes \\
$^{28}$Si$^{32}$S & v=0-0, J=47-46 &  850.801 &  352.365 & 2.02e-13 & no \\
$^{28}$Si$^{32}$S & v=0-0, J=48-47 &  868.795 &  345.067 & 1.98e-13 & no \\
$^{28}$Si$^{32}$S & v=0-0, J=49-48 &  886.781 &  338.068 & 2.04e-13 & yes \\
$^{28}$Si$^{32}$S & v=0-0, J=50-49 &  904.760 &  331.350 & 2.05e-13 & no \\
$^{28}$Si$^{32}$S & v=0-0, J=51-50 &  922.733 &  324.896 & 2.00e-13 & yes \\
$^{28}$Si$^{32}$S & v=0-0, J=52-51 &  940.697 &  318.692 & 2.01e-13 & no \\
$^{28}$Si$^{32}$S & v=0-0, J=53-52 &  958.655 &  312.722 & 2.06e-13 & no \\
$^{28}$Si$^{32}$S & v=0-0, J=54-53 &  976.605 &  306.974 & 2.07e-13 & yes \\
$^{28}$Si$^{32}$S & v=0-0, J=55-54 &  994.547 &  301.436 & 2.05e-13 & no \\
$^{28}$Si$^{32}$S & v=0-0, J=56-55 & 1012.481 &  296.097 & 2.01e-13 & no \\
$^{28}$Si$^{32}$S & v=0-0, J=57-56 & 1030.407 &  290.946 & 2.03e-13 & no \\
$^{28}$Si$^{32}$S & v=0-0, J=58-57 & 1048.325 &  285.973 & 2.07e-13 & no \\
$^{28}$Si$^{32}$S & v=0-0, J=59-58 & 1066.235 &  281.169 & 2.09e-13 & yes \\
$^{28}$Si$^{32}$S & v=0-0, J=60-59 & 1084.136 &  276.527 & 2.09e-13 & no \\
$^{28}$Si$^{32}$S & v=0-0, J=61-60 & 1102.029 &  272.037 & 2.09e-13 & yes \\
$^{28}$Si$^{32}$S & v=0-0, J=62-61 & 1119.913 &  267.693 & 2.08e-13 & yes \\
$^{28}$Si$^{32}$S & v=0-0, J=63-62 & 1137.788 &  263.487 & 2.07e-13 & no \\
$^{28}$Si$^{32}$S & v=0-0, J=64-63 & 1155.654 &  259.414 & 2.06e-13 & yes \\
$^{28}$Si$^{32}$S & v=0-0, J=65-64 & 1173.511 &  255.466 & 2.05e-13 & yes \\
$^{28}$Si$^{32}$S & v=0-0, J=66-65 & 1191.358 &  251.639 & 2.04e-13 & no \\
$^{28}$Si$^{32}$S & v=0-0, J=67-66 & 1209.196 &  247.927 & 2.04e-13 & yes \\
$^{28}$Si$^{32}$S & v=0-0, J=68-67 & 1227.025 &  244.325 & 2.04e-13 & no \\
$^{28}$Si$^{32}$S & v=0-0, J=69-68 & 1244.843 &  240.827 & 2.04e-13 & yes \\
$^{28}$Si$^{32}$S & v=0-0, J=70-69 & 1262.652 &  237.431 & 2.05e-13 & no \\
$^{28}$Si$^{32}$S & v=0-0, J=71-70 & 1280.451 &  234.130 & 2.05e-13 & no \\
$^{28}$Si$^{32}$S & v=0-0, J=72-71 & 1298.240 &  230.922 & 2.06e-13 & yes \\
$^{28}$Si$^{32}$S & v=0-0, J=73-72 & 1316.018 &  227.803 & 2.05e-13 & no \\
$^{28}$Si$^{32}$S & v=0-0, J=74-73 & 1333.786 &  224.768 & 2.04e-13 & yes \\
$^{28}$Si$^{32}$S & v=0-0, J=75-74 & 1351.543 &  221.815 & 2.00e-13 & no \\
$^{28}$Si$^{32}$S & v=0-0, J=76-75 & 1369.290 &  218.940 & 4.81e-14 & yes \\
$^{28}$Si$^{32}$S & v=0-0, J=77-76 & 1387.025 &  216.141 & 1.96e-13 & yes \\
$^{28}$Si$^{32}$S & v=0-0, J=78-77 & 1404.750 &  213.413 & 5.78e-14 & no \\
$^{28}$Si$^{32}$S & v=0-0, J=79-78 & 1422.463 &  210.756 & 1.97e-13 & yes \\
$^{28}$Si$^{32}$S & v=0-0, J=80-79 & 1440.165 &  208.165 & 6.35e-14 & no \\
$^{28}$Si$^{32}$S & v=0-0, J=81-80 & 1457.856 &  205.639 & 6.63e-14 & no \\
$^{28}$Si$^{32}$S & v=0-0, J=82-81 & 1475.535 &  203.175 & 7.07e-14 & no \\
$^{28}$Si$^{32}$S & v=0-0, J=83-82 & 1493.202 &  200.772 & 7.24e-14 & yes \\
$^{28}$Si$^{32}$S & v=0-0, J=84-83 & 1510.858 &  198.425 & 1.83e-13 & yes \\
$^{28}$Si$^{32}$S & v=0-0, J=85-84 & 1528.501 &  196.135 & 7.80e-14 & no \\
$^{28}$Si$^{32}$S & v=0-0, J=86-85 & 1546.132 &  193.898 & 1.79e-13 & no \\
$^{28}$Si$^{32}$S & v=0-0, J=87-86 & 1563.751 &  191.714 & 8.13e-14 & no \\
$^{28}$Si$^{32}$S & v=0-0, J=88-87 & 1581.357 &  189.579 & 1.78e-13 & yes \\
$^{28}$Si$^{32}$S & v=0-0, J=89-88 & 1598.951 &  187.493 & 1.71e-13 & yes \\
$^{28}$Si$^{32}$S & v=0-0, J=90-89 & 1616.532 &  185.454 & 1.72e-13 & no \\
$^{28}$Si$^{32}$S & v=0-0, J=91-90 & 1634.100 &  183.460 & 1.72e-13 & no \\
$^{28}$Si$^{32}$S & v=0-0, J=92-91 & 1651.655 &  181.510 & 1.66e-13 & yes \\
$^{28}$Si$^{32}$S & v=0-0, J=93-92 & 1669.197 &  179.603 & 1.65e-13 & yes \\
$^{28}$Si$^{32}$S & v=0-0, J=94-93 & 1686.725 &  177.736 & 1.60e-13 & yes \\
$^{28}$Si$^{32}$S & v=0-0, J=95-94 & 1704.240 &  175.910 & 1.60e-13 & no \\
$^{28}$Si$^{32}$S & v=0-0, J=96-95 & 1721.742 &  174.122 & 1.55e-13 & no \\
$^{28}$Si$^{32}$S & v=0-0, J=97-96 & 1739.229 &  172.371 & 1.55e-13 & no \\
$^{28}$Si$^{32}$S & v=0-0, J=98-97 & 1756.703 &  170.656 & 1.53e-13 & yes \\
$^{28}$Si$^{32}$S & v=0-0, J=99-98 & 1774.163 &  168.977 & 1.48e-13 & yes \\
$^{28}$Si$^{32}$S & v=0-0, J=100-99 & 1791.608 &  167.331 & 1.44e-13 & yes \\
$^{28}$Si$^{32}$S & v=0-0, J=101-100 & 1809.039 &  165.719 & 1.42e-13 & yes \\
$^{28}$Si$^{32}$S & v=0-0, J=102-101 & 1826.456 &  164.139 & 1.41e-13 & no \\
$^{28}$Si$^{32}$S & v=0-0, J=103-102 & 1843.858 &  162.590 & 1.36e-13 & yes \\
$^{28}$Si$^{32}$S & v=0-0, J=104-103 & 1861.245 &  161.071 & 1.31e-13 & no \\
$^{28}$Si$^{32}$S & v=0-0, J=105-104 & 1878.618 &  159.581 & 1.32e-13 & no \\
$^{28}$Si$^{32}$S & v=0-0, J=106-105 & 1895.975 &  158.120 & 1.27e-13 & yes \\
$^{28}$Si$^{32}$S & v=0-0, J=107-106 & 1913.318 &  156.687 & 1.26e-13 & no \\
$^{28}$Si$^{32}$S & v=0-0, J=108-107 & 1930.644 &  155.281 & 1.19e-13 & yes \\
$^{28}$Si$^{32}$S & v=0-0, J=109-108 & 1947.956 &  153.901 & 1.17e-13 & yes \\
$^{28}$Si$^{32}$S & v=0-0, J=110-109 & 1965.251 &  152.547 & 1.13e-13 & no \\
$^{28}$Si$^{32}$S & v=0-0, J=111-110 & 1982.531 &  151.217 & 1.15e-13 & yes \\
$^{28}$Si$^{32}$S & v=0-0, J=112-111 & 1999.795 &  149.912 & 1.08e-13 & yes \\
$^{28}$Si$^{32}$S & v=0-0, J=113-112 & 2017.043 &  148.630 & 1.06e-13 & no \\
$^{28}$Si$^{32}$S & v=0-0, J=114-113 & 2034.275 &  147.371 & 1.03e-13 & yes \\
$^{28}$Si$^{32}$S & v=0-0, J=115-114 & 2051.491 &  146.134 & 1.03e-13 & no \\
$^{28}$Si$^{32}$S & v=0-0, J=116-115 & 2068.690 &  144.919 & 1.00e-13 & yes \\
$^{28}$Si$^{32}$S & v=0-0, J=117-116 & 2085.872 &  143.725 & 9.45e-14 & no \\
$^{28}$Si$^{32}$S & v=0-0, J=118-117 & 2103.037 &  142.552 & 9.44e-14 & no \\
$^{28}$Si$^{32}$S & v=0-0, J=119-118 & 2120.186 &  141.399 & 9.06e-14 & yes \\
$^{28}$Si$^{32}$S & v=0-0, J=120-119 & 2137.318 &  140.266 & 8.85e-14 & yes \\
$^{28}$Si$^{32}$S & v=0-0, J=121-120 & 2154.432 &  139.151 & 8.56e-14 & yes \\
$^{28}$Si$^{32}$S & v=0-0, J=122-121 & 2171.529 &  138.056 & 8.17e-14 & no \\
$^{28}$Si$^{32}$S & v=0-0, J=123-122 & 2188.609 &  136.979 & 7.89e-14 & yes \\
$^{28}$Si$^{32}$S & v=0-0, J=124-123 & 2205.671 &  135.919 & 7.78e-14 & yes \\
\hline
$^{28}$Si$^{32}$S & v=1-1, J=50-49 &  900.341 &  332.977 & 3.26e-14 & yes \\
$^{28}$Si$^{32}$S & v=1-1, J=51-50 &  918.224 &  326.491 & 3.48e-14 & no \\
$^{28}$Si$^{32}$S & v=1-1, J=52-51 &  936.101 &  320.257 & 3.69e-14 & no \\
$^{28}$Si$^{32}$S & v=1-1, J=53-52 &  953.970 &  314.258 & 3.89e-14 & yes \\
$^{28}$Si$^{32}$S & v=1-1, J=54-53 &  971.831 &  308.482 & 4.08e-14 & yes \\
$^{28}$Si$^{32}$S & v=1-1, J=55-54 &  989.685 &  302.917 & 4.28e-14 & yes \\
$^{28}$Si$^{32}$S & v=1-1, J=56-55 & 1007.530 &  297.552 & 4.47e-14 & no \\
$^{28}$Si$^{32}$S & v=1-1, J=57-56 & 1025.368 &  292.376 & 4.67e-14 & yes \\
$^{28}$Si$^{32}$S & v=1-1, J=58-57 & 1043.197 &  287.378 & 4.86e-14 & no \\
$^{28}$Si$^{32}$S & v=1-1, J=59-58 & 1061.018 &  282.552 & 5.03e-14 & yes \\
$^{28}$Si$^{32}$S & v=1-1, J=60-59 & 1078.831 &  277.886 & 5.21e-14 & yes \\
$^{28}$Si$^{32}$S & v=1-1, J=61-60 & 1096.635 &  273.375 & 5.37e-14 & yes \\
$^{28}$Si$^{32}$S & v=1-1, J=62-61 & 1114.430 &  269.010 & 5.51e-14 & yes \\
$^{28}$Si$^{32}$S & v=1-1, J=63-62 & 1132.217 &  264.784 & 5.61e-14 & yes \\
$^{28}$Si$^{32}$S & v=1-1, J=64-63 & 1149.994 &  260.690 & 5.68e-14 & yes \\
$^{28}$Si$^{32}$S & v=1-1, J=65-64 & 1167.763 &  256.724 & 5.84e-14 & yes \\
$^{28}$Si$^{32}$S & v=1-1, J=66-65 & 1185.522 &  252.878 & 6.09e-14 & yes \\
$^{28}$Si$^{32}$S & v=1-1, J=67-66 & 1203.271 &  249.148 & 6.29e-14 & yes \\
$^{28}$Si$^{32}$S & v=1-1, J=68-67 & 1221.011 &  245.528 & 6.42e-14 & yes \\
$^{28}$Si$^{32}$S & v=1-1, J=69-68 & 1238.741 &  242.014 & 6.46e-14 & yes \\
$^{28}$Si$^{32}$S & v=1-1, J=70-69 & 1256.461 &  238.601 & 6.40e-14 & yes \\
$^{28}$Si$^{32}$S & v=1-1, J=71-70 & 1274.171 &  235.284 & 6.70e-14 & no \\
$^{28}$Si$^{32}$S & v=1-1, J=72-71 & 1291.871 &  232.061 & 6.87e-14 & yes \\
$^{28}$Si$^{32}$S & v=1-1, J=73-72 & 1309.561 &  228.926 & 6.86e-14 & yes \\
$^{28}$Si$^{32}$S & v=1-1, J=74-73 & 1327.240 &  225.877 & 6.86e-14 & yes \\
$^{28}$Si$^{32}$S & v=1-1, J=75-74 & 1344.908 &  222.909 & 7.11e-14 & yes \\
$^{28}$Si$^{32}$S & v=1-1, J=76-75 & 1362.566 &  220.021 & 1.68e-14 & no \\
$^{28}$Si$^{32}$S & v=1-1, J=77-76 & 1380.213 &  217.207 & 1.84e-14 & yes \\
$^{28}$Si$^{32}$S & v=1-1, J=78-77 & 1397.849 &  214.467 & 2.06e-14 & yes \\
$^{28}$Si$^{32}$S & v=1-1, J=79-78 & 1415.473 &  211.797 & 2.16e-14 & yes \\
$^{28}$Si$^{32}$S & v=1-1, J=80-79 & 1433.086 &  209.194 & 7.36e-14 & yes \\
$^{28}$Si$^{32}$S & v=1-1, J=81-80 & 1450.688 &  206.655 & 2.50e-14 & no \\
$^{28}$Si$^{32}$S & v=1-1, J=82-81 & 1468.278 &  204.180 & 2.65e-14 & yes \\
$^{28}$Si$^{32}$S & v=1-1, J=83-82 & 1485.857 &  201.764 & 2.78e-14 & yes \\
$^{28}$Si$^{32}$S & v=1-1, J=84-83 & 1503.423 &  199.407 & 2.93e-14 & yes \\
$^{28}$Si$^{32}$S & v=1-1, J=85-84 & 1520.978 &  197.105 & 7.32e-14 & yes \\
$^{28}$Si$^{32}$S & v=1-1, J=86-85 & 1538.520 &  194.858 & 3.18e-14 & yes \\
$^{28}$Si$^{32}$S & v=1-1, J=87-86 & 1556.050 &  192.662 & 3.18e-14 & no \\
$^{28}$Si$^{32}$S & v=1-1, J=88-87 & 1573.567 &  190.518 & 7.40e-14 & no \\
$^{28}$Si$^{32}$S & v=1-1, J=89-88 & 1591.072 &  188.422 & 7.36e-14 & yes \\
$^{28}$Si$^{32}$S & v=1-1, J=90-89 & 1608.564 &  186.373 & 7.34e-14 & no \\
$^{28}$Si$^{32}$S & v=1-1, J=91-90 & 1626.043 &  184.369 & 7.26e-14 & no \\
$^{28}$Si$^{32}$S & v=1-1, J=92-91 & 1643.509 &  182.410 & 7.13e-14 & yes \\
$^{28}$Si$^{32}$S & v=1-1, J=93-92 & 1660.962 &  180.493 & 7.07e-14 & yes \\
$^{28}$Si$^{32}$S & v=1-1, J=94-93 & 1678.402 &  178.618 & 7.18e-14 & no \\
$^{28}$Si$^{32}$S & v=1-1, J=95-94 & 1695.828 &  176.782 & 6.93e-14 & no \\
$^{28}$Si$^{32}$S & v=1-1, J=96-95 & 1713.240 &  174.986 & 6.97e-14 & no \\
$^{28}$Si$^{32}$S & v=1-1, J=97-96 & 1730.639 &  173.226 & 6.67e-14 & yes \\
$^{28}$Si$^{32}$S & v=1-1, J=98-97 & 1748.024 &  171.504 & 6.84e-14 & no \\
$^{28}$Si$^{32}$S & v=1-1, J=99-98 & 1765.394 &  169.816 & 6.79e-14 & yes \\
$^{28}$Si$^{32}$S & v=1-1, J=100-99 & 1782.751 &  168.163 & 6.66e-14 & yes \\
$^{28}$Si$^{32}$S & v=1-1, J=101-100 & 1800.093 &  166.543 & 6.38e-14 & no \\
$^{28}$Si$^{32}$S & v=1-1, J=102-101 & 1817.420 &  164.955 & 6.52e-14 & no \\
$^{28}$Si$^{32}$S & v=1-1, J=103-102 & 1834.733 &  163.398 & 6.21e-14 & no \\
$^{28}$Si$^{32}$S & v=1-1, J=104-103 & 1852.032 &  161.872 & 6.32e-14 & yes \\
$^{28}$Si$^{32}$S & v=1-1, J=105-104 & 1869.315 &  160.376 & 6.08e-14 & yes \\
$^{28}$Si$^{32}$S & v=1-1, J=106-105 & 1886.583 &  158.908 & 6.00e-14 & yes \\
$^{28}$Si$^{32}$S & v=1-1, J=107-106 & 1903.836 &  157.468 & 6.00e-14 & yes \\
$^{28}$Si$^{32}$S & v=1-1, J=108-107 & 1921.074 &  156.055 & 5.87e-14 & yes \\
$^{28}$Si$^{32}$S & v=1-1, J=109-108 & 1938.296 &  154.668 & 5.77e-14 & yes \\
\hline
\end{longtable}
}% End \longtab

\end{appendix}
\end{document}